\documentclass{article}
\usepackage{graphicx}
\usepackage{amsmath,amssymb}

\begin{document}

\title{Porous composite with negative thermal expansion obtained by photopolymer additive manufacturing}%

\author{Akihiro Takezawa
\thanks{Corresponding author, akihiro@hiroshima-u.ac.jp} \thanks{Division of Mechanical Systems and Applied Mechanics, Institute of Engineering, Hiroshima University, 1-4-1 Kagamiyama, Higashi-Hiroshima, Hiroshima, 739-8527, Japan}
\and
Makoto Kobashi
\thanks{Department of Materials Engineering, Graduate School of Engineering, Nagoya University, Furo-cho, Chikusa-ku, Nagoya 464-8603, Japan}%
\and
Mitsuru Kitamura$^\dag$
}

\maketitle

\section*{Abstract}

Additive manufacturing (AM) could be a novel method of fabricating composite and porous materials having various effective performances based on mechanisms of their internal geometries. Materials fabricated by AM could rapidly be used in industrial application since they could easily be embedded in the target part employing the same AM process used for the bulk material. Furthermore, multi-material AM has greater potential than usual single-material AM in producing materials with effective properties. Negative thermal expansion is a representative effective material property realized by designing a composite made of two materials with different coefficients of thermal expansion. In this study, we developed a porous composite having planar negative thermal expansion by employing multi-material photopolymer AM. After measurement of the physical properties of bulk photopolymers, the internal geometry was designed by topology optimization, which is the most effective structural optimization in terms of both minimizing thermal stress and maximizing stiffness. The designed structure was converted to a three-dimensional STL model, which is a native digital format of AM, and assembled as a test piece. The thermal expansions of the specimens were measured using a laser scanning dilatometer. The test pieces clearly showed negative thermal expansion around room temperature.

\section*{Main Document}
Additive manufacturing (AM) or rapid prototyping technology has received much attention as an innovative manufacturing technique allowing quick fabrication of detailed and complicated objects \cite{gibson2010,frazier2014}. Recent progress of such technology has improved the manufacturing accuracy and level of detail to a composite or porous material scale (about 10$\mu$m) of a three-dimensional (3D) structure. Thus, AM could be applied to the formation of not only macro-scale parts but also new composite or porous materials. Such materials would have various effective properties according to the mechanisms of their internal geometries.

Innovative materials have indeed been developed by integrating mathematical and numerical design methods and AM technologies; e.g., tissue engineering a bone scaffold having compatibility with human bone in terms of stiffness and permeability \cite{hollister2005,castilho2013} and developing materials with a negative Poisson's ratio \cite{schwerdtfeger2011,andreassen2014}. The greatest advantage of the additive manufacture of innovative materials is fast practical realization. In the conventional development of materials, even though innovative techniques have been developed in the laboratory, issues relating to bulk production and machining need to be overcome to allow practical use. In contrast, a geometry-based material fabricated employing commercial AM technology could be directly used in engineering parts. Moreover, functionally graduated characteristics could easily be achieved by spatially varying the internal geometry \cite{zhang2015}.

A further advantage of AM is multi-material fabrication although this function is only realized by photopolymer-type AM using commercially available devices. Using multiple materials in the internal structure, there are more degrees of freedoms of realizable effective material properties than in the single-material case \cite{milton2002,torquato2002}. Effective thermal expansion is a representative example. By forming an internal structure using more than two materials with different positive coefficients of thermal expansion (CTEs) and voids, even negative effective thermal expansion could be achieved. With a rise in temperature, usual thermal expansion occurs within the internal structure of materials. However, the mechanism of the internal structure converts this thermal expansion into an inward deformation of the outer shell and macroscopic negative thermal expansion is then observed.

The studies of such internal geometry-based negative thermal expansion materials had been performed by both theoretically \cite{sigmund1996,lakes1996,lakes2007,hirota2015} and experimentally \cite{qi2004, steeves2007,yamamoto2014}. By choosing two general materials with known stiffness and CTE as base materials, the CTE and stiffness of the composite can be tuned by changing the microstructural shape according to elastic mechanics; compounds having a negative CTE are limited to special cases and the characteristic tuning is an important research topic \cite{barrera2005,takenaka2012}. However, manufacturability is a major issue relating to internal geometry-based negative-CTE materials. First, since the internal structure must be composed of at least two materials having certain shapes, machining and assembly processes are required for each cell. This might be critical in mass production. Moreover, special techniques such as microfabrication by co-extrusion and reduction sintering \cite{qi2004} or microelectromechanical system fabrication \cite{yamamoto2014} are required to form a small-scale ($<10$mm) internal structure.

AM can resolve the above issue since it has the potential to realize multi-material production in almost the same process as that used for a single material by changing the added material according to position. However, only photopolymer-type AM is currently available commercially, and the materials of such AM have a smaller CTE gap than the combination of metals introduced in previous experiments \cite{qi2004, steeves2007,yamamoto2014}. Thus, techniques are required to design an internal geometry that provides effective negative thermal expansion through the combination of materials having a small CTE gap. Numerical structural optimizations could be powerful tools for such difficult structural design. These optimizations could automatically lead to the optimal structure employing numerical structural analysis and optimization techniques. In particular, topology optimization could achieve the fundamental optimization of the target structure including a change in the number of holes \cite{bendsoe2003}. Theoretical studies of internal geometry-based negative thermal expansion based on topology optimization have been conducted \cite{sigmund1996, hirota2015}.

Against the background described above, the present study develops an internal geometry-based negative-thermal-expansion material fabricated by multilateral photopolymer AM. After measuring physical properties of photopolymers, the internal geometry is designed to maximize negative thermal expansion employing numerical topology optimization. The effective physical properties of the internal structure are calculated employing numerical homogenization. Test pieces composed of the designed internal structure are fabricated by photopolymer AM. Negative thermal expansion of the internal structure is then experimentally verified by measuring the thermal deformation of the test pieces using a laser scanning dilatometer.

The present work investigates the thermal expansion phenomena of a composite porous material composed of an internal structure that has a periodic layout in a plane. We assume the thermal expansion behavior of the material in the internal structure follows the constitutive relation
\begin{equation}
\sigma_{ij}=C_{ijkl}(\epsilon_{kl}-\Delta T\alpha_{kl})=C_{ijkl}\beta_{kl},
\label{eq01}
\end{equation}
where $\boldsymbol{\sigma}$, $\bold{C}$, $\boldsymbol{\epsilon}$, $\boldsymbol{\alpha}$ and $\boldsymbol{\beta}$ are the stress tensor, the elastic tensor, the strain tensor, the CTE tensor and the thermal stress tensor and $\Delta T$ is a temperature change from a reference temperature. In this research, we realize negative thermal expansion by considering the effective (or macroscopic, averaging) stiffness and CTE of the porous composite. The macroscopic effective physical properties of the periodic structure can be calculated employing numerical homogenization \cite{guedes1990, sigmund1997jmps}. The effective elastic tensor $\bold{C}^H$, CTE tensor $\boldsymbol{\alpha}^H$ and thermal stress tensor $\boldsymbol{\beta}^H$ of the periodic structure composed of a unit cell $Y$ are calculated as

\begin{gather}
C^H_{ijkl} = \frac{1}{|Y|}\int_Y\left(C_{ijkl}-C_{ijpq}\frac{\partial \chi^{kl}_p}{\partial y_q}\right)\text{d}Y,
\label{eq02}\\
\alpha^H_{ij}=[C^H_{ijkl}]^{-1}\beta^H_{pq}=[C^H_{ijkl}]^{-1}\frac{1}{|Y|}\int_Y\left(\beta_{pq}-C_{pqkl}\frac{\partial \psi_k}{\partial y_l}\right)\text{d}Y,
\label{eq03}
\end{gather}
where $\boldsymbol{\chi}$ and $\boldsymbol{\psi}$ are the displacements obtained by solving the problem of $Y$-periodic cells expressed as
\begin{gather}
\int_YC_{ijpq}\left(\delta_{pk}\delta_{ql}-\frac{\partial \chi^{kl}_p}{\partial y_q}\right)\frac{\partial v_i}{\partial y_j}\text{d}Y=0,
\label{eq04}\\
\int_Y\left(\beta_{ij}-C_{ijkl}\frac{\partial \psi_k}{\partial y_l}\right)\frac{\partial v_i}{\partial y_j}\text{d}Y=0,
\label{eq05}
\end{gather}
where $\bold{v}$ is an arbitrary test function.

The composite porous internal geometry is designed by optimizing the above effective stiffness and CTE. We set a square domain as the base shape of the internal geometry and design it by allocating two types of photopolymers and voids. Solid isotropic material with penalization (SIMP)-based multiphase topology optimization is introduced as the design optimization tool for the internal geometry, which is represented by the layout of three phases, two materials and voids, in the specified domain by defining two artificial density functions $\phi_1$ and $\phi_2$ $(0<\phi_1,\phi_2\le 1)$. The elastic modulus and CTE are formulated as functions of $\phi_1$ and $\phi_2$, which represent the existence of materials and the kinds of materials respectively. The local Young modulus and CTE of the design target domain are represented as

\begin{gather}
E(\phi_1,\phi_2) = \phi_1^3\left\{\phi_2 E_1+(1-\phi_2)E_2\right\},
\label{eq06}\\
\alpha(\phi_2) = \phi_2\alpha_1+(1-\phi_2)\alpha_2,
\label{eq07}
\end{gather}
where $E_1$, $E_2$, $\alpha_1$ and $\alpha_2$ are Young's modulus and the CTE of photopolymers 1 and 2 respectively. In other words, $\phi_1\approx1$ and $\phi_2\approx1$ means that photopolymer 1 exists, $\phi_1\approx1$ and $\phi_2\approx0$ means that photopolymer 2 exists, and $\phi_1\approx0$ means the void exists. Since the state cannot be identified from intermediate values between 0 and 1, it should be avoided by contriving the problem and optimizer settings. To realize the isotropic effective CTE, symmetry is assumed on the center and diagonal lines of the square design domain. Thus, distributions of $\phi_1$ and $\phi_2$ are optimized only on the 1/8 domain shown in Fig. \ref{oc} (a).

When only the effective CTE is considered in designing a structure having negative thermal expansion, a structure having very low stiffness could result, which would be unsuitable for fabrication and experiment. Moreover, the effective thermal stress tensor $\boldsymbol{\beta}^H$ is preferred over the effective CTE $\boldsymbol{\alpha}^H$ both in \eqref{eq03} in terms of the easiness of the calculation of its gradient used in the optimizer. We thus intend to maintain a certain level of effective stiffness while reducing the effective CTE by setting the multi-objective function as

\begin{equation}
\mathop{\text{minimize}} \limits_{\phi_1,\phi_2} J(\phi_1,\phi_2) = -w*C^H_{iiii} + (1-w)*\beta^H_{ii}\ (i=1,2).
\label{eq08}
\end{equation}

The steps to the optimization are as follows. We first solve \eqref{eq04} and \eqref{eq05} employing the finite element method (FEM) and a commercial FEM solver COMSOL Multiphysics (COMSOL Inc., USA). Second, the effective physical properties in \eqref{eq02} and \eqref{eq03} and the objective function in \eqref{eq08} are calculated. Since the density function is updated by gradient-based algorithms, the first-order gradient of the objective function is then calculated using the adjoint method \cite{allaire2007}. The density functions $\phi_1$ and $\phi_2$ are updated by the method of moving asymptotes (MMA) \cite{svanberg1987} in the first stage of the optimization. In the latter stage, to obtain a clear shape avoiding intermediate values, the density function is updated using the phase field method \cite{takezawa2010jcp}.

According to the numerical design methodology described above, the photopolymer composite internal geometry is designed. An Objet Connex 500 (Stratasys Ltd., USA), which is the only commercial photopolymerization manufacturing machine offering multi-material 3D printing, is used to fabricate the test pieces. The machine produces 3D structures by spraying liquid photopolymer onto a build tray in thin layers and exposing the photopolymer to ultraviolet light. In producing a structure from multiple materials, a rigid material, a rubber-like material and several admixture materials can be used.

\begin{figure}[H]
\centering
\includegraphics[scale=1.0,clip]{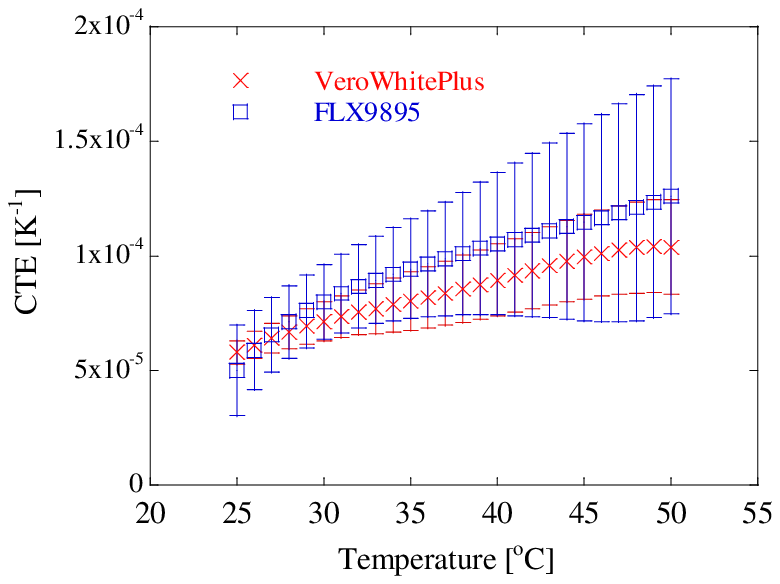}
\caption{Measured CTEs of bulk VeroWhitePlus RGD835 and FLX9895-DM photopolymers. The temperature rising rate is 1$^\circ$C/min. The temperature range is between RT and 50$^\circ$C. $N=3$.}
\label{cte}
\end{figure}

\begin{figure}[H]
\centering
\includegraphics[scale=1.0,clip]{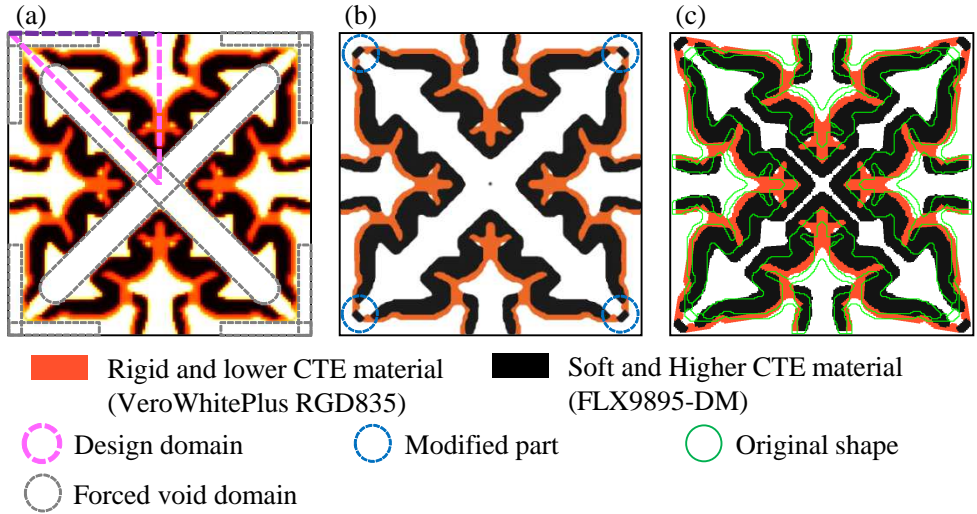}
\caption{Optimal configurations and 2D view of the STL model. (a) Original topology optimization result. Pink dotted lines indicate the optimized domain. Gray dotted lines indicate the forced void domain for maintaining the deformation range. (b) 2D view of the STL model generated from the topology optimization result. The hinge-like part of the stiff material on the corner was replaced with thicker soft material as shown in blue dotted circles. (c) Thermal deformation shape obtained from the reanalysis of the STL model. The green lines indicate the original shape.}
\label{oc}
\end{figure}

We measured Young's modulus by tensile testing in a temperature-controlled bath at 20, 30 and 40$^\circ$C and measured CTE using a connecting rod dilatometer for several materials between RT and 50$^\circ$C, which is close to 45$^\circ$C, the heat deflection temperature (ASTM D648, 0.45MPa) of a rigid material. According to the obtained physical properties, we chose VeroWhitePlus RGD835 and FLX9895-DM, which is an admixture material of VeroWhitePlus and TangoBlackPlus, because this combination achieves a certain level of stiffness and a certain CTE difference in the temperature range between RT and 40$^\circ$C. The measured physical properties are listed in Table \ref{tableE} and plotted in Fig. \ref{cte}. Poisson's ratios of these materials were measured only at 40$^\circ$C and they are both about 0.48. Although CTEs of both materials have strong temperature dependency and dispersion within this temperature range, we found a difference in the CTE on average, which is a necessary condition for a negative effective CTE. With the aim of realizing a negative effective CTE at 40$^\circ$C, which is the temperature at which the CTE difference between the materials is a maximum, we set Young's modulus and the CTE in \eqref{eq06} and \eqref{eq07} as follows according to the measured average physical properties:
\begin{equation}
E_1=290\text{MPa},\ E_2=5.0\text{MPa}, \alpha_1=1.0\times10^{-4}\text{K}^{-1}\ \text{and}\ \alpha_2=1.2\times10^{-4}\text{K}^{-1}.
\label{eq09}
\end{equation}

\begin{table}
\caption{Measured Young's modulus of VeroWhitePlus RGD835 and FLX9895-DM photopolymers. $N=3$ for each temperature.}
\label{tableE}%
\centering
\begin{tabular}{ccccc}
\hline
\hline
& \multicolumn{2}{l}{VeroWhitePlus RGD835} & \multicolumn{2}{l}{FLX9895-DM}\\
Temperature [$^\circ$C]& Average [MPa]& SD [MPa]& Average [MPa]& SD [MPa]\\
\hline
20 & 1996.2 & 111.2 & 67.8 & 2.4\\
30 & 1023.3 & 228.0 & 13.0 & 5.4\\
40 & 288.5 & 34.6 & 3.8 & 0.8\\
\hline
\hline
\end{tabular}
\end{table}

Topology optimization is then performed using the optimal shape shown in Fig. 2(a) of ref\cite{sigmund1996} as the initial shape. We first set the physical properties as, $E_1=290$MPa, $E_2=290$MPa, $\alpha_1=1\times10^{-4}$K$^{-1}$ and $\alpha_2=1\times10^{-3}$K$^{-1}$, which has the same ratio of each $E$ and $\alpha$ with the example used in ref \cite{sigmund1996}. They were varied to the values in \eqref{eq09} gradually during the optimization procedure. Figure \ref{oc} (a) shows the final converged solution. We had introduced the forced void domain (marked by gray dots) in which no material can exist in the final stage of optimization to maintain enough space for thermal deformation.

We then obtained 3D data in STL format by extruding the two-dimensional (2D) optimal shape in the thickness direction. The 2D view of the STL is shown in Fig. \ref{oc} (b). The hinge-like thin rigid photopolymer 1 on the edge of the diagonal seemed not to endure the fabrication and was replaced by the thicker soft photopolymer 2. The effective in-plane elastic tensor and CTE tensor calculated for the 3D model with the physical properties in \eqref{eq09} are presented in Table \ref{tableH}. Negative effective CTEs were certainly obtained. An effective negative Poisson's ratio was also incidentally observed. The free thermal deformation shape of this model is shown in Fig. \ref{oc} (c). The mechanism of the negative thermal expansion is as follows. Bending deformations occur in the bilayer composed of different CTE materials. Since these parts are connected by a hinge-like part, the parts connecting with the next cell deform inward. When the inward deformation of the connecting parts is larger than the axial thermal expansion of the connecting parts, there is macroscopic negative thermal expansion. According to the homogenization analysis for the STL model using Young's modulus shown in Table \ref{tableE} and the parameterized CTE, differences in the CTEs of photopolymers of at least 1.15 times is required at 20, 30 and 40 $^\circ$C to realize the negative effective CTE.

\begin{table}
\caption{Effective planar material properties of the optimal internal structure with the bulk physical properties in \eqref{eq09}}
\label{tableH}%
\centering
\begin{tabular}{cccc}
\hline
\hline
\multicolumn{3}{l}{Effective elastic tensor [MPa]} & \multicolumn{1}{l}{Effective CTE tensor [K$^{-1}$]}\\
$C^H_{1111}(=C^H_{2222})$&$C^H_{1122}(=C^H_{2211})$&$C^H_{1212}$&$\alpha^H_{11}(=\alpha^H_{22})$ \\
\hline
$4.12\times10^{-2}$ & $-1.01\times10^{-2}$ & $1.90\times10^{-3}$ & $-3.53\times10^{-5}$\\
\hline
\hline
\end{tabular}
\end{table}

We employed laser scanning to measure the thermal deformation since it is suitable for relatively large specimens. The testing device was an SL-1600A (Shinagawa Refractories Co., Ltd.), which measures the axial thermal deformation of rod-shaped specimens by laser scanning.

\begin{figure}[H]
\centering
\includegraphics[scale=1.0,clip]{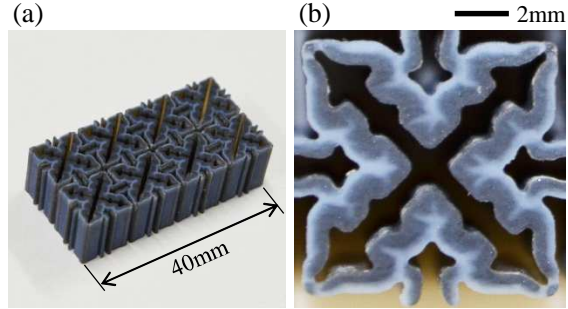}
\caption{Picture of a test piece. (a) Outline view. (b) Close-up view corresponding to the STL model shown in Fig. \ref{oc} (b). }
\label{tp}
\end{figure}

By arranging the eight base STLs shown in Fig. \ref{oc} (b) in two lines, the STL model of the test piece was constructed. The size of each internal structure was $10\times10times10$mm and the total size of the test piece was $40\times20times10$mm. Figure \ref{tp} shows the outline and close-up pictures of the fabricated test piece. From the close-up picture of the internal geometry, it can be said that almost the same shape as the original STL model was achieved except for small branch-like features.

\begin{figure}[H]
\centering
\includegraphics[scale=1.0,clip]{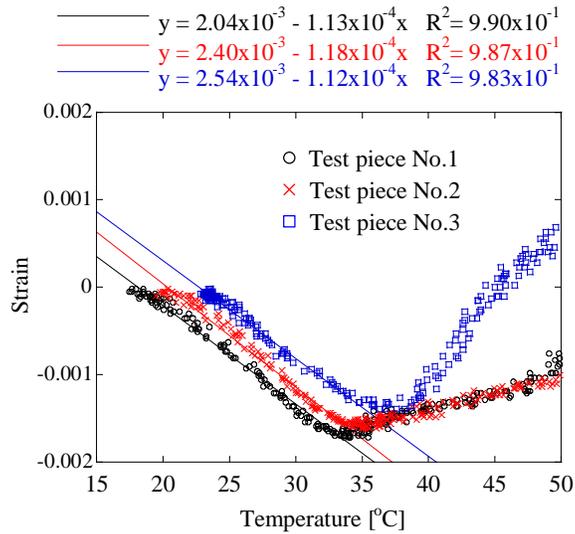}
\caption{Thermal deformation of test pieces measured by a laser scanning dilatometer. The temperature rising rate is 1$^\circ$C/min. The temperature range is between RT and 50$^\circ$C. RTs of measurements No. 1, 2 and 3 are 17.45, 19.45 and 22.85$^\circ$C respectively. Approximation lines are plotted for the data within the temperature range between RT and 34$^\circ$C.}
\label{results}
\end{figure}

The thermal expansions of the long sides of three test pieces were measured in the temperature range between RT and 50$^\circ$C with a temperature rising rate of 1$^\circ$C/min. The measured strains are plotted in Fig. \ref{results}. Negative thermal expansion was clearly observed for each result between RT and about 45$^\circ$C. The early thermal expansions are linear, and corresponding CTEs were calculated as 1.12--1.18K$^{-1}$ between RT and 34$^\circ$C. Since the variation in CTEs is relatively low in such a temperature range according to the bulk CTE results shown in Fig. \ref{cte}, stable differences of CTEs of the two materials are guessed for this linear region. However, beyond approximately 34 to 36$^\circ$, the CTE differences decreased and negative thermal expansion reduced or became positive. Owing to the variability of the physical properties of photopolymers, the performances of the fabricated porous composite also varied widely. However, planar negative thermal expansion characteristics could be clearly designed for photopolymer porous composites.

In summary, we fabricated a porous material with planar negative thermal expansion employing multi-material photopolymer AM. The internal geometry was designed using topology optimization. The mechanism of the effective negative thermal expansion is inward deformation resulting from the combination of bending and hinge mechanisms. The 2D optimal internal geometry was converted to an STL model and assembled as a test piece. The thermal expansion of the test piece was measured with a laser scanning dilatometer. Negative thermal expansion corresponding to less than $-1\times10^{-4}$K$^{-1}$ was certainly observed for each test piece of the $N=3$ experiment. Owing to the temperature dependence and variable physical properties of the photopolymer, the control of the performance of the developed materials appears difficult. However, although multi-material AM is currently commercially available only in the case of a photopolymer, multi-material metal AM is under development \cite{hofmann2014,sahasrabudhe2015}. If such AM is realized, the design process proposed in this paper could be used for materials that are more stiff and stable than photopolymer.

This work was supported by JSPS KAKENHI Grant number 15K12557.


\end{document}